\documentclass[12pt]{article}

\usepackage{graphicx,amssymb,amsmath}
\usepackage{amsthm}
\usepackage{graphicx}
\usepackage{algorithm}
\usepackage{algorithmic}
\newtheorem{lemma}{Lemma}
\newtheorem{theorem}[lemma]{Theorem}
\newtheorem{corrolary}[lemma]{Corollary}

\newcommand\blfootnote[1]{%
  \begingroup
  \renewcommand\thefootnote{}\footnote{#1}%
  \addtocounter{footnote}{-1}%
  \endgroup
}
\begin{document}
\title{Lower Bound for Sculpture Garden Problem}
\author{M. Eskandari
\footnote{ Corresponding author: Faculty of Mathematical  Sciences, Alzahra University, Tehran, Iran. eskandari@alzahra.ac.ir}
, B. Sadeghi Bigham
\footnote{Institute  for Advanced Studies  in Basic Sciences (IASBS), Zanjan, Iran.  b\_sadeghi\_b@iasbs.ac.ir }  }

\maketitle
 \blfootnote{The Authors' name are in alphabetical order.}

\begin{abstract}
The purpose of the current study is to investigate a special case of art gallery problem, namely  Sculpture Garden Problem.
In the said problem, for a given polygon $P$, the ultimate goal is to place the minimum number of guards to define the interior polygon $P$ by applying a monotone Boolean formula composed of the guards.
 As the findings indicate, the conjecture about the issue that in the worst case, $n-2$ guards are required to describe any $n$-gon (Eppstein et al. 2007) can be conclusively proved. 
 
\end{abstract} 

\section{Introduction}
\begin{figure}[ht]
\centering
\includegraphics[width=4in]{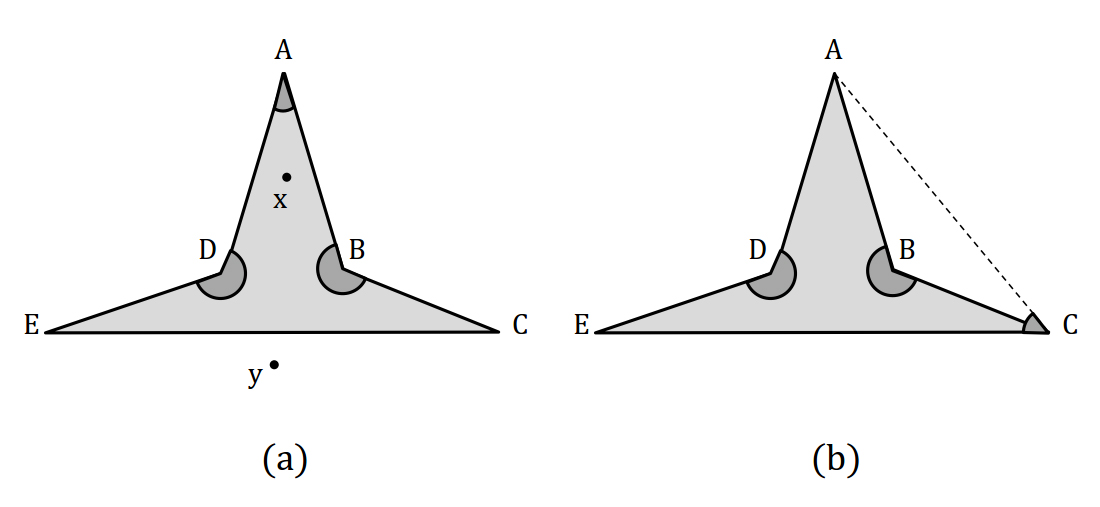}
\caption{ (a) Natural angle guards do not suffice to define the polygon; (b) Coverage by three guards (formula  A.B.D define the polygon) \cite{2} }
\label{Fig1}
\end{figure}

A large and growing body of literature about computational geometry has explored the art gallery  problem.
The main goal in this problem is to place the minimum number of point guards inside a polygon 
$P$ then, the set of guards can see the whole  $P$.
The number of guards that suffices and sometimes necessary for any arbitrary polygon with $n$ vertices is  $\lfloor n/3 \rfloor $ \cite{12}.
The main goal in our study is to find the minimum set of angle guards by which the geometry of the polygon can be defined through two operations $AND(.) $ and $OR(+)$.
An  $angle \; guard$  $g$
views an infinite wedge of the plane (by going through the involved obstacles) and can be defined as a Boolean predicate,
$ B_g(p) $,  which is $True$ for a given point $ p\in P$ if $p$ is inside the view region of $g$, otherwise it is False.
Given a polygon $P$, the aim is to place a set of angle guards on $ P $ in such a way that the monotone Boolean formula $F_P(p)$ is $True$, if and only if $p$ is inside or on the polygon $P$, otherwise it is False: 
$$F_P(p) =
\left\{
	\begin{array}{ll}
		True  & \mbox{if } p \in P \mbox{ or }\mbox p \in \partial P\\
		False & \mbox{otherwise }
	\end{array}
\right.$$
An angle guard vertex placement is considered as \textit{natural} if all the guards of P have the same view of their corresponding vertices  \cite{1}. As Eppstein et al. stated, a polygon $P$ can be demonstrated in a way that a natural angle guard vertex placement cannot fully distinguish between points on the inside and outside of $P$ which implies  $Steiner$-point guards are sometimes necessary  \cite{1}. According to Figure \ref{Fig1}.a,  even the placement of a natural angle at every vertex of the pentagon is not able to distinguish between the points $x$ and $y$  and at least one unnatural guard is needed to define the polygon (Figure  \ref{Fig1}.b). Consequently the polygon is defined by $F = A.B.D$. 

A variety of cases of problem is present in which the location and angel of view of the guards are different.  We focus on a type of the problem that all the guards are placed on the vertices of $P$. It was a conjecture  \cite{2} that in the worst case,  $n-2$ guards are needed to describe $n$-gon. In this paper, we prove that the conjecture is true and present an algorithm to generate a polygon for a given $n$ which needs at least $n-2$ guards.

In the next section, the Sculpture guard problem is introduced and some applications are mentioned. Section 3 provides the main problem and presents an algorithm for generating the $n$-gon which needs exactly $n-2$ guards to be defined. 
Finally, Section 4, presents the findings of the study and also some suggestions for further research.

\section {Sculpture Garden Problem and Applications}
As  mentioned previously, Sculpture Garden Problem can be considered as a special case of Art Gallery problem. 
There are various problems with similarities and differences with Sculpture garden problem. As Sculpture Garden Problem comes up from localization problems in wireless mobile computing, we wish to determine the position of some devices in a geometric environment.

Sculpture Garden Problem could be used in a localization problems in which a wireless device is used to prove that it belongs to a given polygonal environment. In this case, the locators would be simple and   can broadcast information inside a certain angle. In this context, the Boolean predicates could be associated with secret keys. Therefore, each angle guard $g$ could periodically broadcast a secret key 
$K$ in its transmission angle and consequently the devices in range would have knowledge of this key.
The wireless localization problem with natural and vertex guards is  a NP-hard problem  \cite{5,6}.

In another application, namely constructive solid geometry (CSG), we wish to construct a geometric shape from simple combinations of primitive shapes \cite{10}.
Solutions to the Sculpture guard problem can be used to construct an efficient CSG formula that defines a given polygon $P$.
Prison yard problem seeks a set of guards that can simultaneously see both the interior and exterior of a simple polygon, in which  $[n/2]$ guards are sufficient and sometimes necessary (tight bound) \cite{8}.
Another related problem is Floodlight illumination problem in which the vertex angle guards (called floodlights) with angles of 180 should see a simple polygon \cite{11}. 
Likewise, studies have been conducted on the complexity of illuminating wedges with angle-restricted floodlights placed at a fixed set of points \cite{9}. 

Since we are interested in more than simple observing the inside and outside of a polygon, solution to art gallery or prison yard problem would not change into solutions to sculpture garden problem. In other words, it is intended to establish the time when a point is inside a polygon using only the guards as witnesses.

 Indeed, any polygon $P$ can be triangulated and two of angel guards can be used to define each of the resulting $n+2(h-1)$ triangles, where $h$  is the number of holes in  $P$. This would give rise to a concise formula $F$  defining $P$. However, it uses at least $2n+4(h-1)$ angle guards, which is more constant-depth formulas.

\subsection {Upper and Lower Bounds }
The Sculpture Garden Problem has different types due to the different restrictions as guards could  be observed in varied forms including vertices, edges, interior, or exterior of polygon, the SGP can be manifested in different types, as well. However, in each case, finding the upper and lower bound is a  problem that already was investigated.

An angle guard $g$ with angle $\alpha \in( 0,360)$ is a pair $(a,\omega_{\alpha})$ of a point $a$ and an infinitive wedge $\omega_{\alpha}$ of aperture $\alpha$ at apex $a$ and views $\omega_{\alpha}$.
It can be shown as a Boolean predicate $B_g(p)$, in which for a point $p$ in the plane, $B_g(p)$, is $True$, if
$p$ is inside the angle associated with $g$  otherwise it is $False$.
Given a polygon $P$ with $n$ vertices, we intend to allocate the minimum number of angle guards with arbitrary angles at vertices of $P$. Thus, a monotone Boolean formula, $F_P(.)$,  based on the angle guard predicates, $B_g(.)$ is obtained as follows:
\begin{center}
\begin{equation} 
F_P(p) =
\left\{
	\begin{array}{ll}
		True  &  \forall p\in int(P) \\
		False & \mbox{otherwise }
	\end{array}
\right.
\end{equation}
\end{center}

It is worth mentioning that $int(P)$ is interior of polygon $P$.
If $F_P(p)$ is a solution of sculpture garden problem for a given $P$, $P$ is defined by $F_P$.
The complement of an angle guard $g = (a ,\omega_{\alpha})$ is an angle guard
$g^\prime$ at the point $a$ with angle $2\pi - \alpha$. Hence, the wedge associated with $g^\prime$
is the \textit{complement} of $\omega_{\alpha}$ in plane. If formula $F$ is a solution for sculpture garden problem for polygon $P$, then the complement of  $F$ which is denoted by $F^\prime$ defines the exterior of $P$. To obtain $F^\prime$, initially, we replace every angle guard $g$ by its complement, (i.e. $g^\prime$), and then swap the operations $AND$ and $OR$. 
In addition, we define , a \textit{pocket} of a simple polygon as the areas outside of the polygon and inside of its convex hull.

As Eppstein et al. \cite{1,2} reported, for any polygon $P$, a set of $n+2(h-1)$ angle guards and an associated concise formula $F$ are present solving the Sculpture Garden problem where $h$ is the number of holes in $P$. So, a simple polygon can be defined with $n-2$ guards. 
They have conjectured a class of simple $n$-gons that require at least 
$n-2$ natural vertex guards. The main goal of this paper is to solve this open problem. 
They showed at least $\lceil n/2 \rceil $ 
guards are required to solve the Sculpture Garden problem for any polygon with no two edges lying on a same line.  Besides, for any convex polygon $P$, a natural angel guard vertex placement  is present whose $\lceil n/2 \rceil$ guards are sufficient. 
They showed that $\lceil n/2 \rceil + O(1)$ angle guards suffice to solve the Sculpture Garden problem for pseudo-triangles. Moreover, for any orthogonal polygon $P$ (which is, probably the most likely real world applications)  a set of $ [3(n-2)/4] $
angle guards and an associated concise formula
$F$ are available to solve the Sculpture Garden problem
using  $\lceil n/2 \rceil$
natural angle guards. 
They gave an example of a class of simple polygons containing Sculpture Garden solutions are used
$O (\sqrt{n})$ guards. Afterwards, they  showed that the bound is optimal.
On the contrary, some varied results are obtained for vertex guards. 
As Damian et al. demonstrated \cite{4}
 a class of simple $n$-gons are presented that require at least 
$\lfloor 2n/3 \rfloor -1$ 
guards placed at polygon vertices for localization. 
Through revealing the point that the maximum number of guards to describe any simple polygon on $n$ vertices is roughly observable at $ (3/5 n ,4/5 n)$, Hoffman et al. enhanced both upper and lower bounds for the general setting \cite{7}. 
 Eskandari et al. \cite{3} improved the large upper bound
$n+2(h-1)$ for an arbitrary $n$-gon.
with $h$ holes for placing guards and obtained a tight bound
$(n-\lceil c/2 \rceil -h)$, where $c$ is the number of vertices of convex hull of $P$.
So in simple polygons, this bound is $n-\lceil c/2 \rceil$ which is tight too.
 To complete the first column of Table \ref{tab1}, a new class of polygons entitled  \textit{helix} is introduced in the next section.

\begin{table}
\caption{Number of guards needed for a simple polygon with $n$ vertices.}
\label{tab1} 
\centering
\begin{tabular}{|c|c|c|c|c|}
\hline 
\-\-\- & 
$ Natural\; Vertex $ &
$ Natural $ \-\- &
$ Vertex $ &
$ General $\\
\hline
$ Upper Bound $&
$ Not\; known $ &
$ n-2 \cite{7}$ & $ n-2 \cite{1}$ & $ \lfloor
\frac{ (4n-2)}{5}\rfloor \cite{7} $ \\
\hline
$ Lower Bound $&
$ Not\; known $ &
$ n-2 \cite{7}$ & $ \lfloor \frac{ (2n)}{3} \rfloor -1 \cite{4}$ & $ \lceil \frac{ (3n-4)}{5} \rceil \cite{7}$ \\
\hline
\end{tabular}
\end{table}

\section{Helix Polygon}
In this section, we explore the special class of sculpture garden problem, where, the guards are natural. We manifest the point that the lower bound  for the problem is $n-2$.
To do so, we commence with introducing a class of polygons demanding the exact number of $n-2$  natural guards to be defined. In the next section, we introduce this class of polygons named {\it helix} (See Figure 2).

\subsection{Construction Helix}
An $n$-gon helix polygon (i.e. $H_n$) is constructed by an incremental method using an $n-1$-gon helix, $H_{n-1}$. 
A helix with three vertices is a triangle. By adding two new edges to $H_{n-1}$ and also removing an edge of $H_{n-1}$ , $H_{n}$ is constructed on the basis of $H_{n-1}$ where $n \ge 4$.
 The details are presented in Algorithm 1 and are illustrated in Figure 3. 

 \begin{algorithm}[H]
 \hspace*{\algorithmicindent} \textbf{Input:} Integer number $n\geq 3$ as the number of vertices. \\
  \hspace*{\algorithmicindent} \textbf{Output:} The helix polygon $H_{n}$.
    
   \begin{algorithmic}[1]
   \fontsize{12pt}{13pt}\selectfont
   \STATE Construct $H_{3} = \triangle v_{1} v_{2} v_{3}$, which is an equilateral triangle where $v_2v_3$ is horizontal and the vertices are in counterclockwise order.
   \STATE Choose a positive real number $\delta$ so $0< \delta < \frac{|v_2v_3|}{2 \lfloor \frac{n-1}{3} \rfloor}$
   \STATE $p_1=v_1; p_2=v_2; p_3=v_3$.
   \FOR {$i=4$; $i \leq n$; $i++$} 
   		\STATE $q_1=p_1$; $p_1=$ a point on $v_1v_2 $ such that $|p_1q_1|=\delta$, $a=l_{13}(p_1)$;
   		\STATE $q_2=p_2$; $p_2=$ a point on $v_2v_3 $ such that $|p_2q_2|=\delta$, $b=l_{12}(p_2)$;
   		\STATE $q_3=p_3$; $p_3=$ a point on $v_1v_3 $ such that $|p_3q_3|=\delta$, $c=l_{23}(p_3)$;
   		\IF{$i==4$}   
   			\STATE $l=b$;
   			\ELSE 
   			     \STATE $l=l_{24}(v_{i-2})$;
   		\ENDIF
   		\IF{$i==5$}   
   		   	\STATE $l^{'}=c$;
   		   	\ELSE 
				\STATE $l^{'}=l_{35}(v_{i-2})$;
   		\ENDIF
   		\IF{$i==3k$}  
   		   	\STATE $v_i=a \cap c$;
   		\ENDIF
   		\IF{$i==3k+1$}  
   		   		   	\STATE $v_i=c \cap l$;
   		\ENDIF
		\IF{$i==3k+2$}  
   		   	\STATE $v_i=b \cap l^{'}$;
   		\ENDIF
   		\STATE Add edges $v_iv_{i-1}$ and  $v_iv_{i-2}$.
 		\STATE Remove $v_{i-1}v_{i-2}$ to obtain $H_i$
   \ENDFOR
   \STATE Return $H_n$.
   \end{algorithmic}
   \caption{Constructing Helix Polygon}
   \label{alghelix}
   \end{algorithm}
   
   It is worth noting that the length of a line segment $pq$ is denoted by $|pq|$ and for an arbitrary point $p$, $l_{ij}(p)$ denotes a line parallel  to $v_iv_j$ which passes through $p$.
   
\begin{figure}[ht]
\centering
\includegraphics[width=5in]{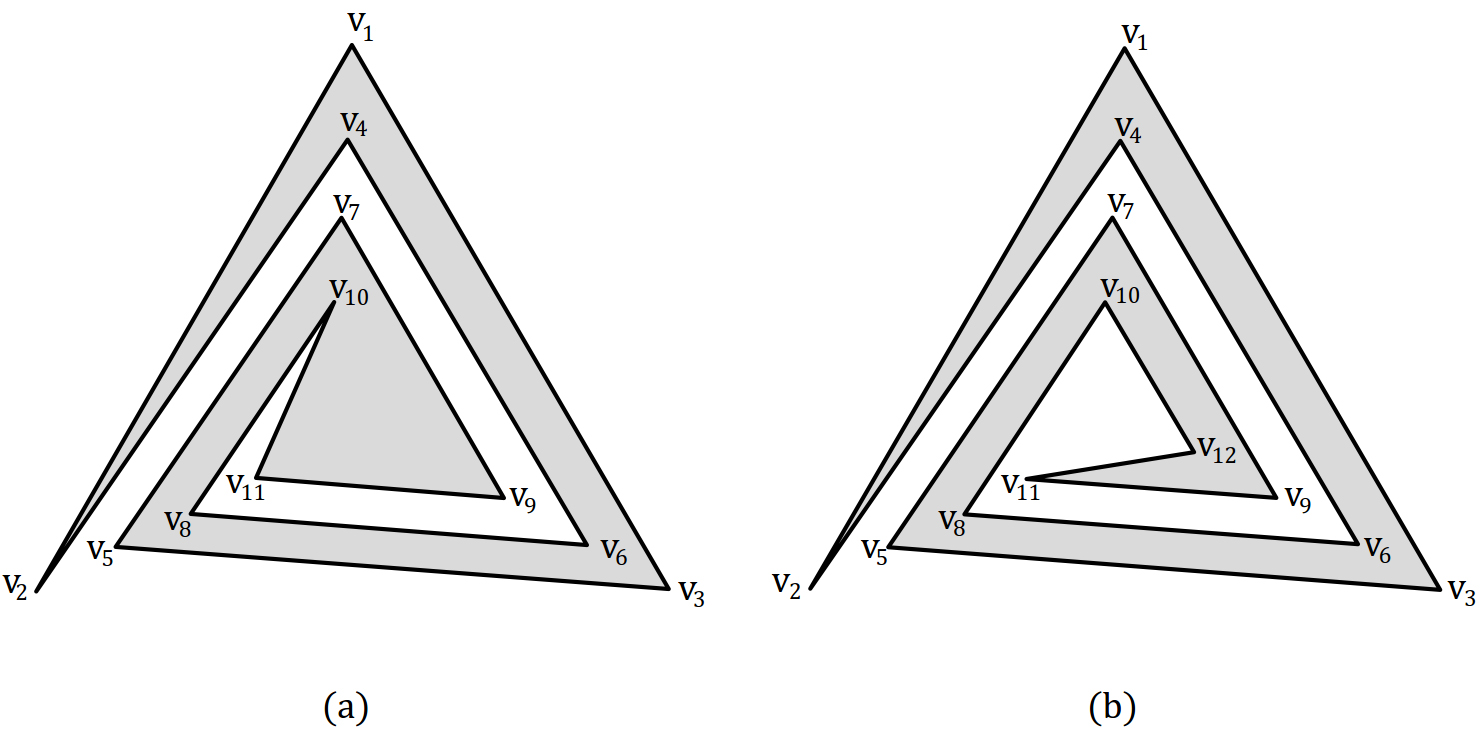}
\caption{(a) A helix with 11 vertices. (b) A helix with 12 vertices. }
\label{Fig2}
\end{figure}

\begin{figure}[ht]
\centering
\includegraphics[width=6in]{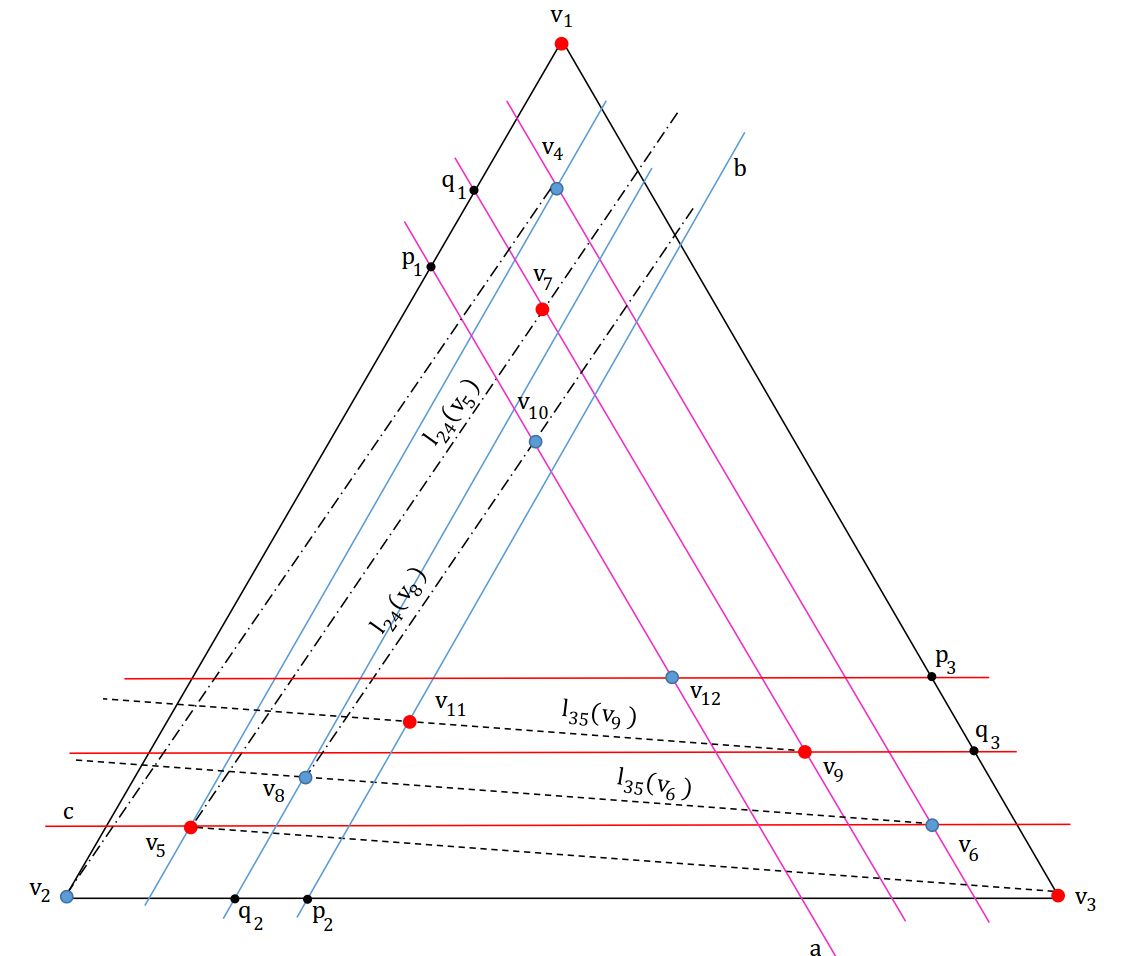}
\caption{Final step of Algorithm 1 for generating $H_{12}$}
\label{Fig3}
\end{figure}

\subsection{Properties of  Helix}
Constructing the helix polygon sheds light that  for $i>2$, vertices $v_{2i}$ are concave and for $i>0$ , vertices $v_{2i +1}$  and $v_2$ are convex (Figure 2).
 
In fact, the pocket of a polygon $P$ is defined as $CH(P)-P$ where $CH(P)$
 is the convex hull of the vertices of $P$.
The pocket of helix polygon with $n$ vertices is a helix polygon with $n-1$ vertices (Figure 4).
The pocket of a polygon $H_n$ is denoted by $P(H_n)$.  For $i$,  $1 \leq i \leq n-1$, the vertices of $P(H_n)$ are called $v^{'}_i$  located on $v_{i+1}$.
For $n>4$, the angle $\widehat {v^{'}_i}$ in $P(H_n)$ is obtained as follows:

\begin{center}
\begin{equation}
\widehat{v^{'}_i}=
\left\{
	\begin{array}{lll}
		\widehat {v_1 v_2 v_3}  - \widehat {v_1 v_2 v_4} & \mbox{ }i=1\\
		\widehat {v_1 v_3 v_2}  - \widehat {v_1 v_3 v_5} & \mbox{ }i=2 \\
		2 \pi - \textrm{interior angle of }  v_{i+1} \textrm{in}\; H_n  & \mbox{ } i\geq 3.
    \end{array}
\right.
\end{equation}
\end{center}

\begin{figure}[ht]
\centering
\includegraphics[width=4in]{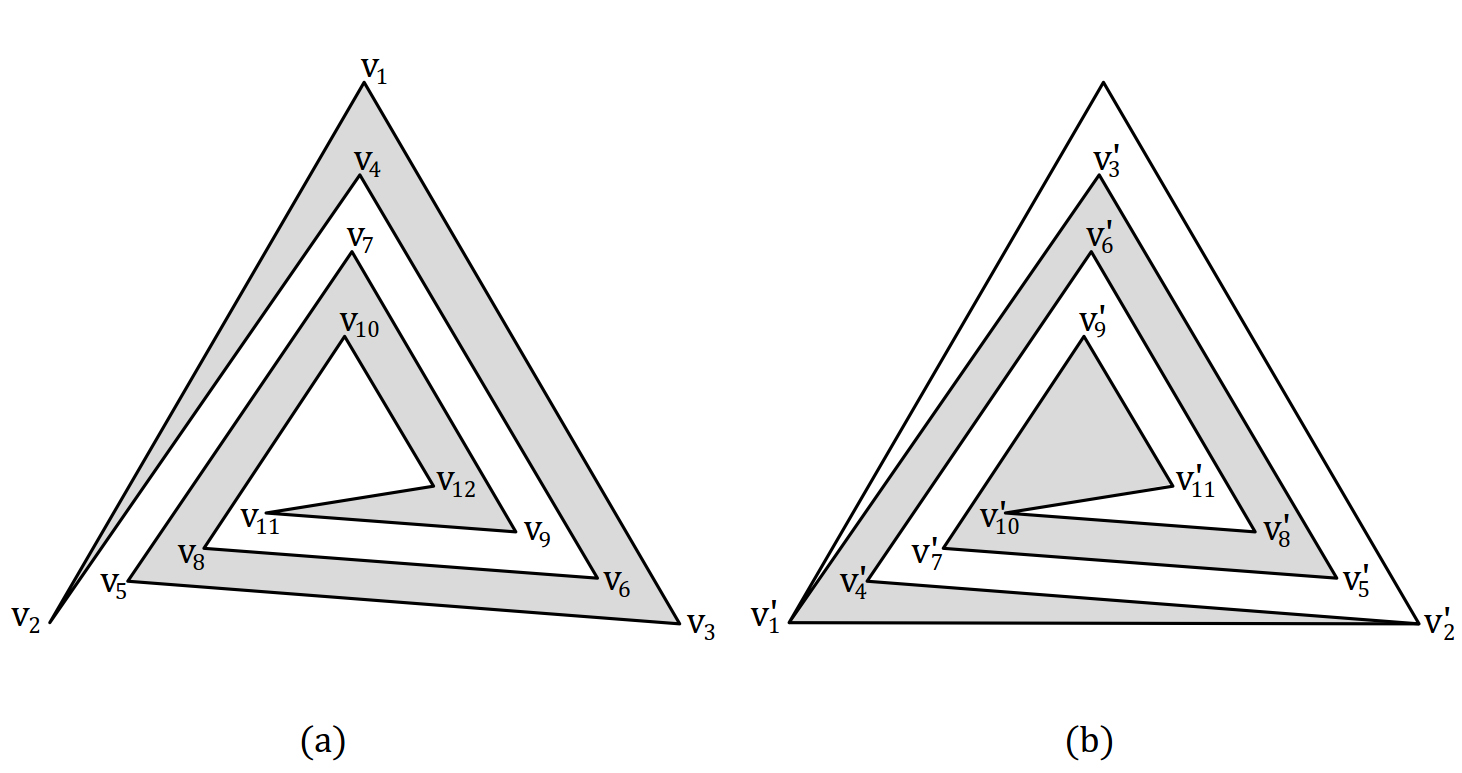}
\caption{(a) Helix $H_{12}$. (b) Pocket of $H_{12}$ which is a helix with 11 vertices. }
\label{Fig4}
\end{figure}

We will show that polygon $H_n$ can be defined by $n-2$ natural vertices guards  which are located on $v_1, v_2, ..., v_{n-3}, v_n$. The Boolean formula, $F_n$, is as below: 
\begin{center}
\begin{equation}
F_n =  \sum_{i=1}^{\lceil n/2\rceil-2} A_i . g_{2i} +A_{[n/2]-1 }. g_n \; 
\end{equation} 
\end{center}
 where 
$A_1= g_1$, $A_{i+1}= A_i .g_{2i+1}$ for all $1 \leq i \leq [ n/2]-2$
 and $g_i$  is the natural vertex guard located on the vertex $v_i$  of $H_n$. 
To clarify this point, $F_8$ can be written as follows:
 \begin{center}
 \begin{equation} F_8= \sum_{i=1}^{2}A_i.g_{2i}+A_5.g_8
 \end{equation} 
 \end{center} 
 where, $A_1=g_1, A_2=g_1.g_3$ and $A_3=g_1.g_3.g_5$. Thus, we have: $$F_8=g_1.g_2+g_1.g_3.g_4+g_1.g_3.g_5.g_8$$
 
 Generally, we expand $F_n$ as follows:
  \begin{center}
  \begin{equation}
  F_n=
 \left\{
 	\begin{array}{lll}
 		\sum_{i=1}^{k-1}(\prod_{j=0}^{i-1}g_{2j+1}).g_{2i} +(\prod_{j=0}^{k-2}g_{2j+1}).g_{n} & \mbox{ }n=2k+1, k \in \mathbb{N} \\
 		\sum_{i=1}^{k-1}(\prod_{j=0}^{i-1}g_{2j+1}).g_{2i} +(\prod_{j=0}^{k-1}g_{2j+1}).g_{n} & \mbox{ }n=2k+2, k \in \mathbb{N}.
 		 
     \end{array}
 \right.
 \end{equation}
 \end{center}
According to Lemma \ref{Lemma_1}, $F_n$ defined by Equation 5 describes $H_n$.

\begin{lemma}
\label{Lemma_1}
Helix polygon $H_n$ can be defined by $n-2$ natural vertex guards $g_i$ ($1 \leq i \leq n; i \neq n-1, n-2$) located on $v_1, v_2, ... v_{n-3}, v_n$ and the correspondent Boolean formula is Formula 5. 
 
\end{lemma}
\begin{proof}
We will prove the lemma by induction. When $n=3$,  $k=1$ and $F_3=g_1.g_3$  clearly defines triangle $H_3$. For $n=4$, $k=1$ and $F_4=g_1.g_4$  defines $H_4$ and it implies that the
Lemma 1 holds for $n=3,4$. Now, for $n \ge 5$, without loss of generality, assume that  $n=2k+2, k \in \mathbb{N}$.
According to Property 2, $P(H_n)$ is a helix polygon with $2k+1$ vertices. By induction hypothesis, $P(H_n)$ can be defined as follows: 
\begin{center}
\begin{equation}
F'=\sum_{i=1}^{k-1}(\prod_{j=0}^{i-1}g'_{2j+1}).g'_{2i} +(\prod_{j=0}^{k-2}g'_{2j+1}).g'_{2k+1}
\end{equation}
\end{center}
 where $g'_i$ is a natural guard on the vertex $v'_i$ of $P(H_n)$. According to correspondence between the vertices of $H_n$ and $P(H_n)$, we have:
\begin{center}
\begin{equation}
g'_i=
 \left\{
 	\begin{array}{lll}
 		g^c_{i+1} & \mbox{ }i \ge 3 \\
 		G_{i+1}.g^c_{i+1} & \mbox{ }i=1,2.
 		 
     \end{array}
 \right.
 \end{equation}
\end{center}
in which $g^c$ is the complement of guard $g$ and $G_2$ and $G_3$ are the guards located on $v_2$ and $v_3$ with the angles $\widehat {v_1 v_2 v_3}$ and $\widehat {v_1 v_3 v_2}$, respectively. So we have:
$$F'=g'_1.g'_2+\sum_{i=2}^{k-1}(\prod_{j=0}^{i-1}g'_{2j+1}).g'_{2i} +(\prod_{j=0}^{k-2}g'_{2j+1}).g'_{2k+1} $$
$$=g'_1.g'_2+\sum_{i=2}^{k-1}g'_1.(\prod_{j=1}^{i-1}g'_{2j+1}).g'_{2i} +g'_1.(\prod_{j=1}^{k-2}g'_{2j+1}).g'_{2k+1} $$

$$=G_2.G_3.g^c_2.g^c_3+\sum_{i=2}^{k-1}G_2.g^c_2.(\prod_{j=1}^{i-1}g^c_{2j+2}).g^c_{2i+1} +G_2.g^c_2.(\prod_{j=1}^{k-2}g^c_{2j+2}).g^c_{2k+2} $$
$$=G_2.[G_3.g^c_2.g^c_3+\sum_{i=2}^{k-1}(\prod_{j=0}^{i-1}g^c_{2j+2}).g^c_{2i+1} +(\prod_{j=0}^{k-2}g^c_{2j+2}).g^c_{2k+2}] $$
$$\implies {(F')}^c=G^c_2+(G^c_3+g_2+g_3).(\prod_{i=2}^{k-1}(\sum_{j=0}^{i-1}g_{2j+2})+g_{2i+1}).(\sum_{j=0}^{k-2}g_{2j+2}+g_{2k+2}) $$
Consider the point that $F=(g_1.g_2).{(F')}^c+(g_1.g_3).(F')^c$. By replacing ${(F')}^c$ from the above relation, we obtain:

$$F=g_1.(g_2+g_2.g_3).(\prod_{i=2}^{k-1}(\sum_{j=0}^{i-1}g_{2j+2})+g_{2i+1}).(\sum_{j=0}^{k-2}g_{2j+2}+g_{2k+2}) $$
$$+g_1.(g_2.g_3+g_3).(\prod_{i=2}^{k-1}(\sum_{j=0}^{i-1}g_{2j+2})+g_{2i+1}).(\sum_{j=0}^{k-2}g_{2j+2}+g_{2k+2}) $$

Note that $g_1.g_2.G^c_2=g_1.g_3.G^c_2=g_1.g_2.G^c_3=g_1.g_3.G^c_3=\varnothing$. So we have:

$$F=(\prod_{i=2}^{k-1}(\sum_{j=0}^{i-1}g_{2j+2})+g_{2i+1}).(\sum_{j=0}^{k-2}(g_{2j+2}+g_{2k+2})).(g_2+g_3).g_1 $$

Now, we show that $F$ can define $H_n$ which contains exactly the natural guards $g_1, g_2, ..., g_{n-3}, g_n$ and can be written in the form of $F_n$.

First, consider the definition of $F$ which contains only natural guards. To prove that $H_n$ can be defined by $F$, let $x$ be an arbitrary point inside $H_n$. We have $g_1(x)=True$ and $(F')^c(x)=True$ ( $x \in H_n \implies  x \notin P(H_n) \implies F'(x)=False \implies (F')^c(x)=True$ ).
There are two cases:
\begin{itemize}
\item{$g_2(x)=True \implies F(x)=(g_1(x).g_2(x)).(F')^c(x)+(g_1(x).g_3(x)).(F')^c(x)=True$}
\item{$g_2(x)=False \implies g_3(x)=True \implies (g_1(x).g_3(x)).(F')^c(x)=True \implies F(x)=True. $}
\end{itemize}

Thus, $F$ can distinguish the interior of $H_n$. Now, let $x \notin H_n$. There are two cases:
\begin{itemize}
\item{$x \in P(H_n) \implies F'(x)=True , (F')^c(x)=False \implies F(x)=False$}
\item{$x \notin P(H_n) \implies x \in Ext(\triangle v_1v_2v_3) \implies g_1(x)=False \implies F(x)=False. $}
\end{itemize}
So, $F$ can distinguish the exterior of $H_n$ as well. Now, we show that $F$ can be written in the form of $F_n$. Let 
$$S=(g_2+g_3).(\prod_{i=2}^{k-1}(\sum_{j=0}^{i-1}g_{2j+2})+g_{2i+1}).(\sum_{j=0}^{k-2}g_{2j+2}+g_{2k+2})$$
and $$T=\sum_{i=1}^{k-1}(\prod_{j=1}^{i-1}(g_{2j+1}).g_{2i})+(\prod_{j=1}^{k-1}(g_{2j+1}).g_{2k+2})$$

Note that $F_n=g_1.T$ and $F=g_1.S$. To prove $F=F_n$, it is sufficient to show that $T=S$. For all integers $r$ where $1 \leq r \leq k-1$, we define $S^{(r)}_i$ as follows:

\begin{center}
\begin{equation}
S^{(r)}_i=
 \left\{
 	\begin{array}{lll}
 		g_{2i+1}+\sum_{j=r}^{i-1}g_{2j+2} & \mbox{ }r \leq i \leq k-1 \\
 		g_{2k+2}+\sum_{j=r}^{k-2}g_{2j+2} & \mbox{ }i=k.
 		 
     \end{array}
 \right.
 \end{equation}
\end{center}
 So, we have: 
\begin{center}
 \begin{equation}
 S^{(1)}_i=
  \left\{
  	\begin{array}{lll}
  		g_{2i+1}+\sum_{j=1}^{i-1}g_{2j+2} & \mbox{ }1 \leq i \leq k-1 \\
  		g_{2k+2}+\sum_{j=1}^{k-2}g_{2j+2} & \mbox{ }i=k.
  		 
      \end{array}
  \right.
  \end{equation}
\end{center}

By definition of $S^{(r)}_i$,  it is clear that:
$$ S=\prod_{i=1}^{k}(g_{2}+S^{(1)}_i)= g_{2}+\prod_{i=1}^{k}S^{(1)}_i$$
On the other hand, $S^{(r)}_i-S^{(r+1)}_i=g_{2r+2}$. Therefore we have: 
$$\prod_{i=r}^{k}S^{(r)}_i=S^{(r)}_r.\prod_{i=r+1}^{k}S^{(r)}_i$$
$$=g_{2r+1}.\prod_{i=r+1}^{k}S^{(r)}_i=g_{2r+1}.\prod_{i=r+1}^{k}(g_{2r+2}+S^{(r+1)}_i)=g_{2r+1}.(g_{2r+2}+\prod_{i=r+1}^{k}S^{(r+1)}_i)$$
By applying obtained recursive relation, $k-2$ times on $S=g_2+\prod_{i=1}^{k}S^{(1)}_i$, $S=T$.  In this respect, we have:
$$S=g_2+\prod_{i=1}^{k}S^{(1)}_i=g_2+g_3.(g_4+\prod_{i=2}^{k}S^{(2)}_i)=g_2+g_3.g_4+g_3.\prod_{i=2}^{k}S^{(2)}_i$$
$$S=g_2+g_3.g_4+g_3.g_5.g_6+g_3.g_5.\prod_{i=3}^{k}S^{(3)}_i$$
After $t$ times, we have:
$$S=(\sum_{i=1}^{t+1}(\prod_{j=1}^{i-1}g_{2j+1}).g_{2i})+\prod_{j=1}^{t}g_{2j+1}.\prod_{i=t+1}^{k}S^{(t+1)}_i$$
So after $k-2$ times, we have:
$$S=(\sum_{i=1}^{k-1}(\prod_{j=1}^{i-1}g_{2j+1}).g_{2i})+\prod_{j=1}^{k-2}g_{2j+1}.\prod_{i=k-1}^{k}S^{(k-1)}_i$$
Note that $\prod_{i=k-1}^{k}S^{(k-1)}_i=S^{(k-1)}_{k-1}.S^{(k-1)}_{k}=g_{2k-1}.g_{2k+2}$, therefore, 
$$S=(\sum_{i=1}^{k-1}(\prod_{j=1}^{i-1}g_{2j+1}).g_{2i})+\prod_{j=1}^{k-1}g_{2j+1}.g_{2k+2}=T.$$
$S=T$ implies $F=F_n$ which means that $F$ can be written in the form of $F_n$ and could define $H_n$.
\end{proof}

\subsection{Necessity of $n-2$  Natural Vertex Guards for Helix}
In this section, we will prove that it is impossible to define helix polygon with  fewer than $n-2$ natural vertex guards. 

\begin{lemma}
\label{g1}
Every arbitrary set of natural vertex guards $G$ which defines $H_n$ contains  $g_1$, a natural vertex guard on $v_1$. The final formula is in the form of $F= F_1.g_1$, where $F_1$ is a Boolean expression of $G-\{g_1\}$. 
\end{lemma}
\begin{proof}
 Let $G$ is an arbitrary set of natural guards which defines $H_n$ by Boolean formula $F$. Suppose for a contradiction that $g_1$ does not belong to $G$. Since $v_1v_2$ and $v_1v_3$ are edges of $H_n$, $G$ should contain two natural guards on $v_2$ and $v_3$ which are called $g_2$ and $g_3$, respectively. So $F$ can be written in the general form $F=g_2.g_3.T_1+g_2.T_2+g_3.T_3$ where $T_i$s are Boolean formulas which do not contain $g_1$, $g_2$ and $g_3$. This will result in a contradiction, mentioned below.
 
 Consider two regions $R_1$ and $R_2$ as shown in Figure 5. Let $x$ be an arbitrary point inside $R_1$ or $R_2$. So we have: $g_2(x)=True$ and $g_3(x)=False$. So,
  \begin{center}
  \begin{equation}
  \label{star}
  F(x)= T_2(x)
   \end{equation}
 \end{center}
 
 Also, note that we can expand $T_2$ in the following general form:
   \begin{center}
   \begin{equation}
   \label{2star}
   T_2=T^{(1)}_2+T^{(2)}_2+...+T^{(l)}_2
    \end{equation}
  \end{center}
 where $T_2^{(i)}$s are the multiplication of some natural guards in $G$. Let $x \in R_1$ be an arbitrary point. So from Equation \ref{star}, it is implied that:
 $$x \in R_1 \implies T_2(x)=F(x)=True$$
 
 Therefore, at least one of the expressions of $T_2$ should be True. Without loss of generality, it can be called $T^{(1)}_2$ and is written as $T^{(1)}_2=g_{i_1}.g_{i_2}. \cdots .g_{i_m}$ 
where $g_{i_j}$ is the natural guards of $G$ and $i_j=1,2,3$. Since $T^{(1)}_2(x)=True$, we have: 
 $$\forall j: 1\leq j \leq m: g_{i_j}(x)=True$$
  Regarding the structure of $H_n$, none of indices $i_j$ can be odd. This is because we know that 
 $$\forall i \geq 1:  g_{2i+1}(x)=False$$
 Now, let $y \in R_2$ be an arbitrary point. Due to the structure of $H_n$, it is inferred that for all $\forall y \in R_2$, we have:
 $$\forall i \geq 2: g_{2i}(y)=True$$.
 So, $T^{(1)}_2(x)=True$ and from Equation \ref{2star},  $T_2(y)=True$ is obtained and consequently  $F(y)=T_2(y)=True$ (due to Equation \ref{star}). Nevertheless, $y \notin H_n$ which is a contradiction.
 With regard to existence of $g_1$ in $G$, $F$ can be written in the form of $F=g_1.T_1+T_2$ where $T_2$ does not contain $g_1$. Indeed, $g_1.T_1+g_1.T_2$  defines $H_n$ as well. Let $x\in H_n$, then
$g_1(x)=True$ and  $g_1(x).T_1(x)+g_1(x).T_2(x)=g_1(x).T_1(x)+T_2(x)=F(x)=True$. If $y\notin H_n$ and
$g_1(y)=True$, then $g_1(y).T_1(y)+g_1(y).T_2(y)=g_1(y).T_1(y)+T_2(y)=F(y)=False$. On the other hand, if $g_1(y)=False$, then $g_1(y).T_1(y)+g_1(y).T_2(y)=False$. So $F=g_1.(T_1+T_2)$ defines $H_n$. In other words, $F$ can be
expressed as $F=g_1.F_1$.
\begin{figure}[ht]
\centering
\includegraphics[width=4in]{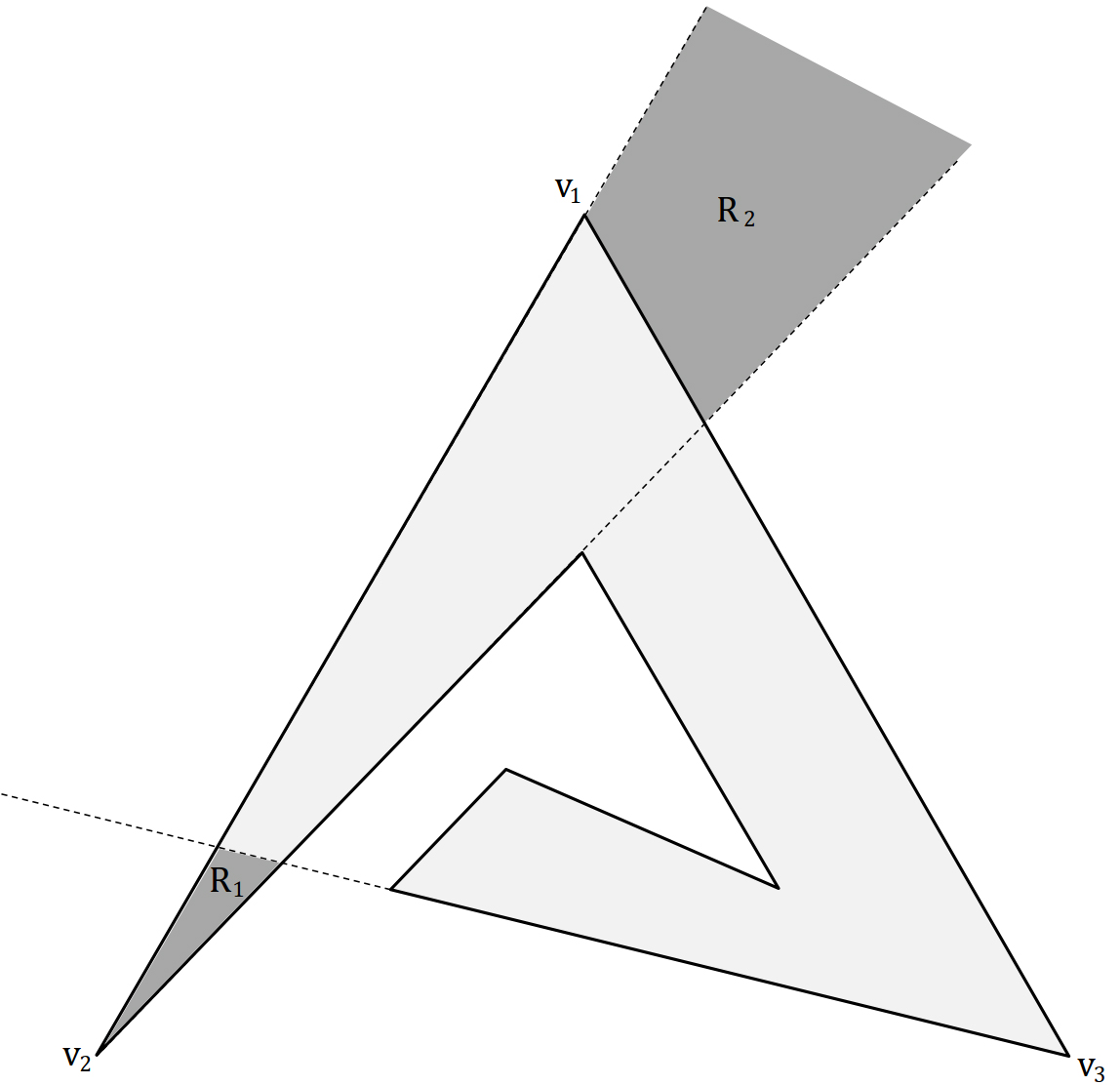}
\caption{Regions $R_1$ and $R_2$ can be distinguishable without existence of $g_1$ in the formula.}
\label{Fig5}
\end{figure}

\end{proof}
\begin{lemma}
\label{g2}
Every arbitrary set of natural vertex guards $G$ defining $H_n$ contains  $g_2$ (i.e. a natural guard on $v_2$). The final formula is in the form of $F= g_1.(g_2+F_2)$, where $F_2$ is a Boolean expression of $G-\{g_1, g_2\}$. 
\end{lemma}
\begin{proof}
Let $G$ is an arbitrary set of natural guards which  defines $H_n$ by Boolean formula $F$, for $n \geq 4$. Suppose, a contradiction in which $g_2$ does not belong to $G$. From Lemma \ref{g1}, we can write $$F=g_1.(T_1+T_2+...+T_l)$$ where $T_i$s are Boolean expression of natural guards of $G$.

Consider two regions $R_1$ and $R_3$ as shown in Figure 6. Let $x \in R_1$ is an arbitrary point. We have $F(x)=g_1(x).(T_1(x)+T_2(x)+...+T_l(x))=True$. Thus, at least one of the expressions $T_i$s is True.
 Without loss of generality, it can be named $T_1$ and is  written as $T_1=g_{i_1}.g_{i_2}....g_{i_m}$ where $g_{i_j}$s are some natural guards in $G$ and $i_j \neq 1,2$. Since $T_1(x)=True$, we have 
 $$\forall j,  1 \leq j \leq m: g_{i_j}(x)=True$$
Regarding the structure of $H_n$, $i_j$s are even, because $\forall i \geq 2: g_{2i}(x)=True$.

Now, let $y \in R_3$ be an arbitrary point. Obviously, $\forall i \geq 2$, $g_{2i}(y)=True$ which implies $T_1(y)=True$. Then $F(y)=True$. However, $ y \notin H_n$ which denotes a contradiction.

In addition, $F$ can be manifested as below.
$$F=g_1.(g_2.T_1+T_2)$$
Now note that for all points which are located in the interior (or exterior) of $H_n$, the above formula has the same value with the formula $g_1.(g_2+T_2)$.
This fact can be shown easily by considering all cases. Then $F=g_1.(g_2+F_1)$.

\begin{figure}[ht]
\centering
\includegraphics[width=4in]{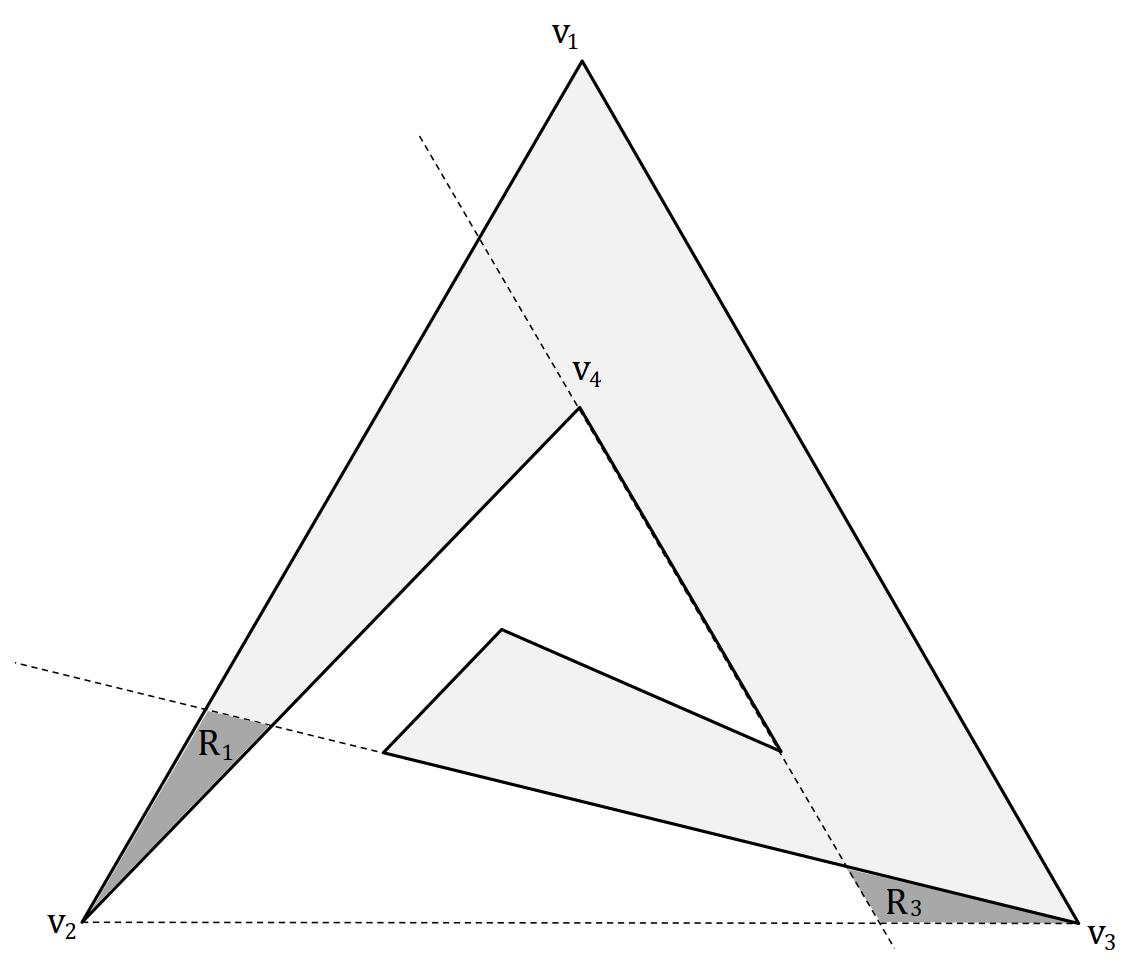}
\caption{Regions $R_1$ and $R_3$ can be distinguishable without existence of $g_2$ in the formula.}
\label{Fig6}
\end{figure} 
\end{proof}
\begin{corrolary}
It is not possible to define $H_5$ with less than 3 natural vertex guards. The formula  is $F= g_1.( g_2 +g_5)$.
\end{corrolary}
\begin{proof}
Regarding Lemmas 2 and 3, $F$ can be written as $g_1.(g_2+F_2)$. The edges $v_4v_5$ and $v_3v_5$ should have at least one guard on their endpoints. The optimal possibility is to locate a guard on
$v_5$ as their intersection point. Clearly, $g_1.(g_2.g_5)$ defines $H_5$.

\end{proof}
\begin{lemma}
\label{g3}
Every arbitrary set of natural vertex guards $G$ which defines $H_n$ contains  $g_3$ which is a natural guard on $v_3$. The final formula is in the form of $F= g_1.(g_2+g_3. F_3)$ where $F_3$ is a Boolean expression of $G-\{g_1, g_2, g_3\}$. 
\end{lemma}
\begin{proof}
Let $G$ be an arbitrary set of natural guards which defines $H_n$ by Boolean formula $F$ for $n \geq 6$. Suppose  contradiction in which $g_3$ does not belong to $G$. By Lemma 3, $F=g_1.(g_2+F_2)$. 
Assume that $F_2=T_1+T_2+...+T_l$ where $T_i$s are multiplication of natural guards in $G$. 

Consider two regions $R_3$ and $R_4$ as shown in Figure 7.  Let $x \in R_4$ be an arbitrary point. We have $F(x) =F_2(x)=True$. So at least one of $T_i$s is True.
Without loss of generality, we call it $T_1$ which can be expressed as follows:
$$T_1=g_{i_1}.g_{i_2}....g_{i_m}$$ where $g_{i_j}$s are natural guards in $G$ and $i_j \neq 1,2,3$.

Since $T_1(x)=True$, we have:
$$\forall j, 1 \leq j \leq m: g_{i_j}(x)=True$$
Therefore $i_j$s are even. Now, let $y \in R_3$ be an arbitrary point. This point casts light that for all $i \geq 2$, $g_{2i}(y)=True$ which implies $T_1(y)=True$. Then, $F(y)=True$. However, $y \notin H_n$ showing a contradiction. So, $g_3 \in G$.

In addition, $F$ can be written as bellow:
$$F+g_1.(g_2+g_3.F_3)$$

It can be easily obtained from equivalency of $g_1.(g_2+F_2)$ and $g_1.(g_2+g_3.F_3)$ for all points with respect to $H_n$.

\begin{figure}[ht]
\centering
\includegraphics[width=4in]{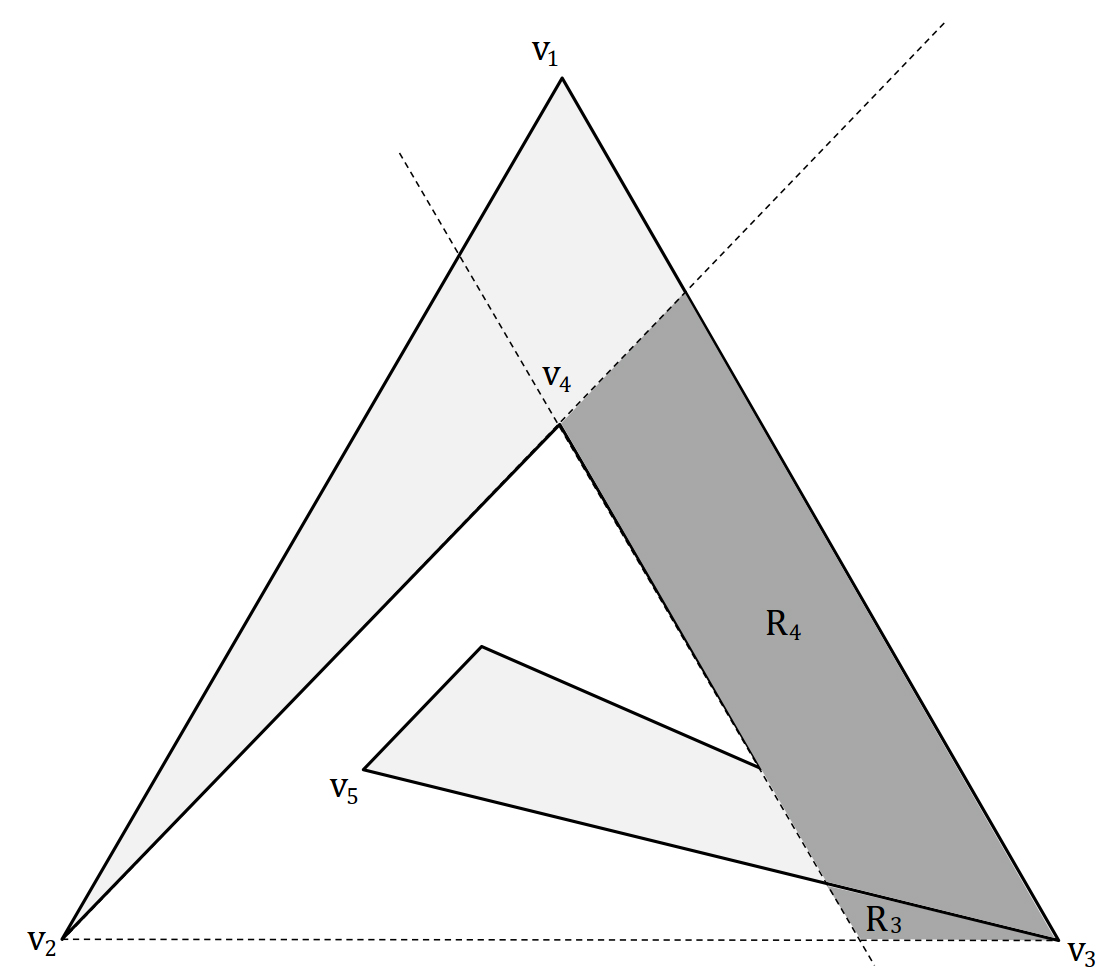}
\caption{Regions $R_3$ and $R_4$ can be distinguishable without existence of $g_3$ in the formula.}
\label{Fig7}
\end{figure}

\end{proof}
\begin{corrolary}
It is not possible to define $H_5$ with fewer than 3 natural vertex guards. The formula is $F=g_1.(g_2+ g_3.g_6)$.
\end{corrolary}
\begin{proof}
Considering Lemma 5, $H_6$ can be defined by $F=g_1.(g_2+g_3.F_3)$. Since $v_4v_6$ and $v_6v_5$ are two edges of $H_6$, it is required to place at least one guard on one of the endpoints of these two edges.
The optimal placement is to place a guard on $v_6$. Obviously, $H_6$ can be defined by  $F=g_1.(g_2+ g_3.g_6)$.
\end{proof}

\begin{lemma}
\label{g4}
Every arbitrary set of natural vertex guards $G$ which defines $H_n$, for all $n\geq 7$, contains  $g_4$, a natural guard on $v_4$. The final formula is in the form of $F= g_1.(g_2+g_3. (g_4 + F_4))$ where $F_4$ is a Boolean expression of $G-\{g_1, g_2, g_3, g_4\}$. 
\end{lemma}
\begin{proof}
Let $G$ be an arbitrary set of natural guards defining $H_n$ by Boolean formula $F$ and $ n \geq 7$. Suppose  a contradiction in which $g_4$ does not belong to $G$.
By Lemma 5, $F= g_1.(g_2+g_3. F_3)$. It is intended to show that $F_3=g_4+F_4$. Assume that $F_3=T_1+T_2+...+T_l$ where $T_i$s are multiplication of natural guards of $G$.
Consider two regions $R_5$ and $R_6$  shown in Figure 8. Let $x \in R_5$ be an arbitrary point, then  $F(x)=F_3(x)=True$. So, at least one of $T_i$s is True. 
Without loss of generality, t is called $T_1$ which can be expressed as bellow:
$$T_1=g_{i_1}.g_{i_2}....g_{i_m}$$ where $g_{i_j}$s are natural guards and $i_j \neq 1,2,3,4$. Since $T_1(x)=True$, for all $j: 1 \leq j \leq m$, $g_{i_j}(x)=True$.
Therefore $i_j$s are even. Now let $y \in R_6$ e an arbitrary point (see Figure 8).  It is clear that for all $i \geq 2$, $g_{2i}(y)=True$ which implies $T_1(y)=True$. Then, $F(y)=True$. However, $y \notin H_n$ indicating a contradiction. So $g_4 \in G$. 
In addition, $F$ can be written as bellow:
$$F= g_1.(g_2+g_3.(g_4 + F_4))$$
It can be easily shown that $g_1.(g_2+g_3.F_3)$ is equivalent with $g_1.(g_2+g_3.(g_4 + F_4))$ for all the points inside or outside of $H_n$.

 \begin{figure}[ht]
\centering
\includegraphics[width=4in]{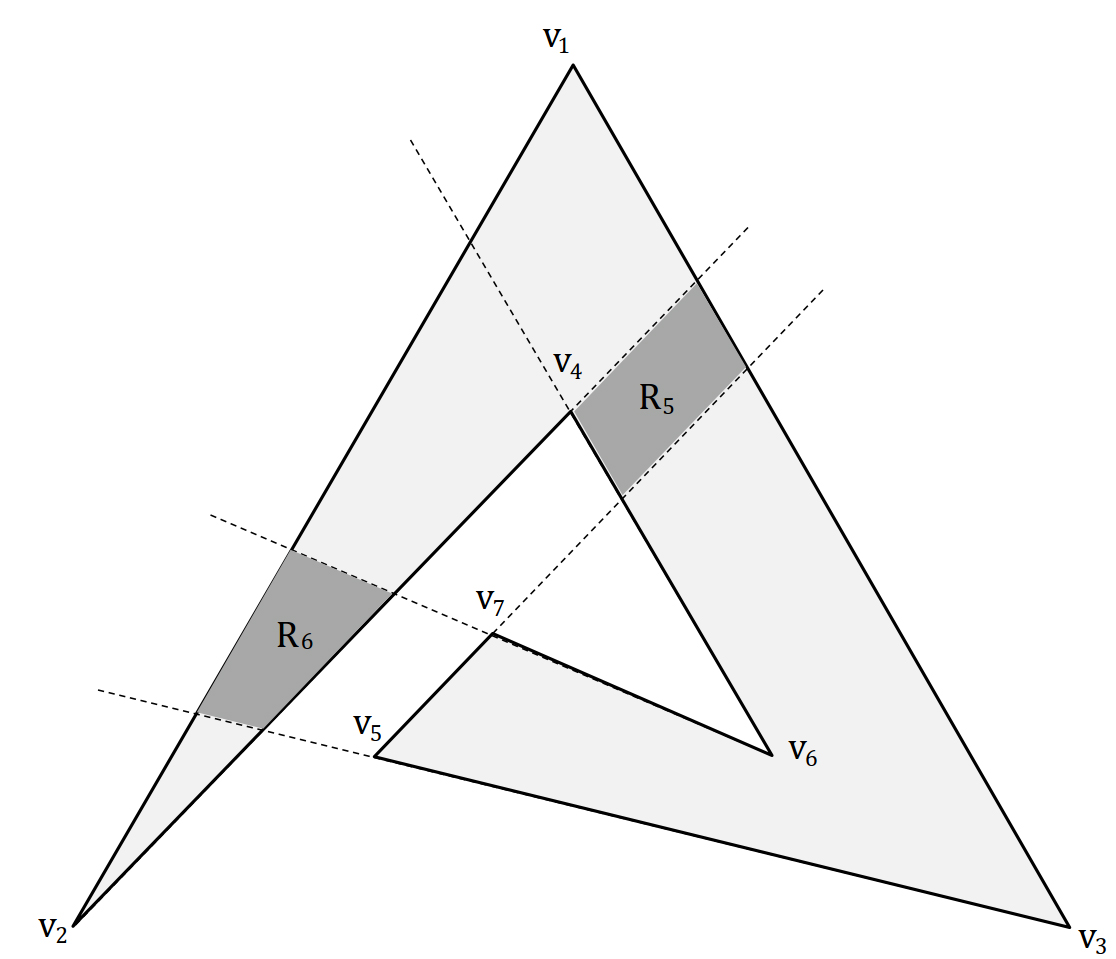}
\caption{Regions $R_5$ and $R_6$ can be distinguishable without existence of $g_4$ in the formula.}
\label{Fig8}
\end{figure}

\end{proof}
\begin{lemma}
Let $H_n$ be defined by $F=g_1.(g_2+g_3. (g_4 + F_4))$. Formula $F_P=g_1^{'} .(g_2^{'}+g_3^{'}. F_4^{c})$ defines the pocket of $H_n$, $P(H_n)$ where $g_i^{'}$s are defined in Equation 7.
\end{lemma}

\begin{proof}
Let $x \in P(H_n)$ and $y \notin P(H_n)$ be two arbitrary points (see Figure 9). Then, it can be demonstrated  that $F_P(x)=True$ and $F_P(y)=False$. Also,  $g_2^{'}=G_3.g_3^{c}$ and $g_3^{'}=g_4^{c}$, so: 
$x \in P(H_n) \implies g_1^{'}(x)=True$ and $G_3(x)=True$. So
 \begin{center}
   \begin{equation}
   \label{starcorollary8}
   F_P(x)=g_3^{c}(x)+ g_4^{c}(x).F_4^{c}(x)
    \end{equation}
  \end{center}
On the other hand, $x \in P(H_n)$ implies that $F(x)=False$ and $g_1(x)=True$. Hence,
$$F(x)=g_2(x)+g_3(x).(g_4(x)+F_4(x))=False \implies g_3(x).(g_4(x)+F_4(x))=False$$
, so 
 \begin{center}
   \begin{equation}
   \label{2starcorollary8}
   g_3^{c}(x)+ g_4^{c}(x).F_4^{c}(x)=True
    \end{equation}
  \end{center}
  
From Equations \ref{starcorollary8} and \ref{2starcorollary8},  $F_P(x)=True$ can be obtained. 

 If $g^{'}(y)=False$, for $y \notin P(H_n)$,  $F_P(y)=False$ and the process of our calculations has been completed. Assume that  $g^{'}(y)=True$, then as $y \notin P(H_n)$, consequently $g_2^{'}(y)=False$ (see Figure 9). If $g_3^{'}(y)=False$,  $F_P(y)=False$.
 Now, suppose that $g_3^{'}(y)=True$ (i.e. $g_4(y)=False$). This assumption implies that $y \in H_n$ and we have  
$g_1(y)=True$, $g_2(y)=False$, $g_31(y)=True$ and $g_4(y)=False$.
Since $y \in H_n$, $F_(y)=F_4(y)$ and consequently, $F_4(y) =True$. So, we have $F_P(y)=False$. This means that $F_P$ defines $P(H_n)$.

 \begin{figure}[ht]
\centering
\includegraphics[width=4in]{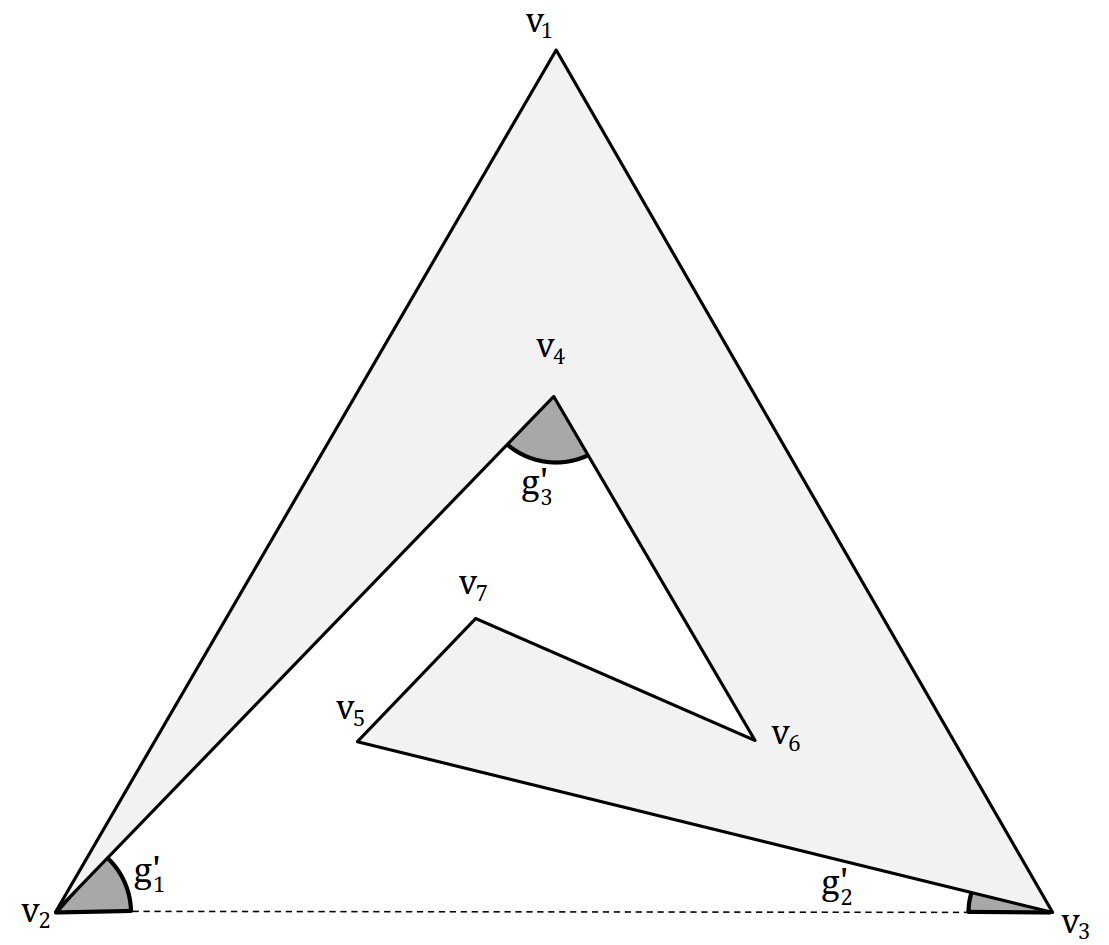}
\caption{Natural vertex guards for packet of helix}
\label{Fig9}
\end{figure}

\end{proof}		   

\begin{theorem}
$H_n$ requires at least $n-2$ natural vertex guards.
\end{theorem}

\begin{proof}
We  prove this theorem by induction. It is clear that for $n=4$,  $H_4$ is a tetragon and cannot be defined by fewer than two guards. We proved this in corollaries 4 and 6 for $n-5$.
Now, assume that this holds for $n-1$, and we have to prove it for $n$ where $n \geq 7$. 
Let $F$ be a Boolean formula which defines $H_n$. With regard to Lemma 7, $F=g_1.(g_2+(g_3. (g_4+F_4))$. Let $m$ be the number of natural guards used in $F$.
From Lemma 8, $P(H_n)$, a helix with $n-1$ vertices, can be defined by $F_P=g_1^{'} .(g_2^{'}+g_3^{'}. F_4^{c})$ which contains $m-1$ guards. By induction hypothesis, $P(H_n)$ cannot be defined by fewer than $(n-1)-2$ natural guards. So $m-1$ cannot be less than $n-3$ and hence $m$ cannot be fewer than $n-2$.
\end{proof}

\begin{theorem}
$H_n$ requires exactly $n-2$ natural vertex guards.
\end{theorem}
\begin{proof}
From Lemma 1 and Theorem 9, it is obviously implied that $H_n$ requires exactly $n-2$ natural vertex guards.
\end{proof}

As we proved, there is an $n$-gon which needs exactly $n-2$ natural vertex guards to be defined. This implies that $n-2$ is the lower bound. 

\section{Conclusion:} 
Epstein et al. \cite{2} in 2007  conjectured that for a given number $n$,  at least one simple polygon is present that requires 
   $ n-2$  natural vertex  guards to describe the polygon.  Introducing a new class of polygons named helix polygon,  we proved the conjecture. Further research should be done to investigate the bounds for special cases of polygons (e.g. orthogonal polygons).
\section*{Acknowledgements}
We deeply thank our colleagues in RoboCG (Robotics and Computational Geometry) lab for their insight and expertise that greatly assisted the research.
 We also appreciate  Arya Falahi for the efforts he made to solve our Latex problems.

\bibliographystyle{unsrt}
\bibliography{Reference}

\begin{thebibliography}{10}

\bibitem{12}
Mark De~Berg, Marc Van~Kreveld, Mark Overmars, and Otfried~Cheong Schwarzkopf.
\newblock {\em Computational geometry}.
\newblock Springer, 2000.

\bibitem{1}
David Eppstein, Michael~T Goodrich, and Nodari Sitchinava.
\newblock Guard placement for efficient point-in-polygon proofs.
\newblock In {\em Proceedings of the twenty-third annual symposium on
  Computational geometry}, pages 27--36. ACM, 2007.

\bibitem{2}
David Eppstein, Michael~T Goodrich, and Nodari Sitchinava.
\newblock Guard placement for wireless localization.
\newblock {\em arXiv preprint cs/0603057}, 2006.

\bibitem{5}
Tobias Christ, Michael Hoffmann, and Yoshio Okamoto.
\newblock Natural wireless localization is np-hard.
\newblock In {\em Abstracts 25th European Workshop Comput. Geom}, pages
  175--178. Citeseer, 2009.

\bibitem{6}
Tobias Christ and Michael Hoffmann.
\newblock Wireless localization with vertex guards is np-hard.
\newblock In {\em CCCG}, pages 149--152. Citeseer, 2009.

\bibitem{10}
David Dobkin, Leonidas Guibas, John Hershberger, and Jack Snoeyink.
\newblock {\em An efficient algorithm for finding the CSG representation of a
  simple polygon}, volume~22.
\newblock ACM, 1988.

\bibitem{8}
Zolt{\'a}n F{\"u}redi and Daniel~J. Kleitman.
\newblock The prison yard problem.
\newblock {\em Combinatorica}, 14(3):287--300, 1994.

\bibitem{11}
Vladimir Estivill-Castro, Joseph O'Rourke, Jorge Urrutia, and Dianna Xu.
\newblock Illumination of polygons with vertex lights.
\newblock {\em Information Processing Letters}, 56(1):9--13, 1995.

\bibitem{9}
William Steiger and Ileana Streinu.
\newblock Illumination by floodlights.
\newblock {\em Computational Geometry}, 10(1):57--70, 1998.

\bibitem{4}
Mirela Damian, Robin Flatland, Joseph O'Rourke, and Suneeta Ramaswami.
\newblock A new lower bound on guard placement for wireless localization.
\newblock {\em arXiv preprint arXiv:0709.3554}, 2007.

\bibitem{7}
Tobias Christ, Michael Hoffmann, Yoshio Okamoto, and Takeaki Uno.
\newblock Improved bounds for wireless localization.
\newblock In {\em Algorithm Theory--SWAT 2008}, pages 77--89. Springer, 2008.

\bibitem{3}
M~Eskandari, A~Mohades, and B~Sadeghi Bigham.
\newblock On the number of guards in sculpture garden problem.
\newblock {\em World Applied Sciences Journal}, 10(10):1255--1263, 2010.

\end{thebibliography}
\end{document}